%% file: main.tex
\newif\ifSingleColumn
\SingleColumnfalse
\ifSingleColumn
\documentclass[draftclsnofoot, onecolumn, 12pt]{IEEEtran}
\else
\documentclass[conference]{IEEEtran}
\fi
\IEEEoverridecommandlockouts
\usepackage{cite}
\usepackage{amsmath,amssymb,amsfonts}
\usepackage{graphicx}
\usepackage{textcomp}
\usepackage{xcolor}
\usepackage{subcaption}
\usepackage{comment}
\usepackage{bm}
\usepackage{float}
\usepackage[noend]{algpseudocode}
\usepackage{algorithmicx,algorithm}
\usepackage[top=0.75in, bottom=0.8in, left=0.625in, right=0.625in]{geometry}
\newcommand{\figwidth}{0.33\linewidth}
\renewcommand{\thetable}{\arabic{table}}

\input{defines.tex}

\newcommand{\beq}{\begin{equation}}
\newcommand{\eeq}{\end{equation}}

\def\BibTeX{{\rm B\kern-.05em{\sc i\kern-.025em b}\kern-.08em
    T\kern-.1667em\lower.7ex\hbox{E}\kern-.125emX}}
\begin{document}

\title{\linespread{1.2}\huge{Enabling High Error Tolerance in Satellite Video Transmissions by Generative Semantic Communication}}

\author{\linespread{1.25}
\IEEEauthorblockN{
\normalsize{Zixin~Zhao}\IEEEauthorrefmark{1},
\normalsize{Jingzhi~Hu}\IEEEauthorrefmark{2}, and
\normalsize{Geoffrey Ye Li}\IEEEauthorrefmark{2}\\
}
\thanks{© 2026 IEEE. Personal use of this material is permitted. Permission from IEEE must be obtained for all other uses, in any current or future media, including reprinting/republishing this material for advertising or promotional purposes, creating new collective works, for resale or redistribution to servers or lists, or reuse of any copyrighted component of this work in other works.}
\thanks{Work was done when Z. Zhao was a visiting student at Imperial College London, UK. Jingzhi Hu is the corresponding author.} 
\IEEEauthorblockA{
	\IEEEauthorrefmark{1}\small{School of Electronics and Information Technology, Sun Yat-Sen University, Guangzhou, China,}\\
	\IEEEauthorrefmark{2}\small{Department of Electrical and Electronic Engineering, Imperial College London, London, UK.}\\}
}

\maketitle

\input{sections/1_abstract}
\ifSingleColumn
\newpage
\fi

\input{sections/2_introduction}

\input{sections/3_systemModel}

\input{sections/4_probFormulation}

\input{sections/5_algDesign}

\input{sections/6_evaluation}

\input{sections/7_conclusion}

\bibliographystyle{IEEEtran}
\bibliography{bibilio}

\end{document}

%% file: defines.tex
\newcommand{\rS}{{\mathrm{S}}}
\newcommand{\rL}{{\mathrm{L}}}
\newcommand{\rt}{{\mathrm{t}}}
\newcommand{\rT}{{\mathrm{T}}}
\newcommand{\rtr}{{\mathrm{tr}}}
\newcommand{\rs}{{\mathrm{s}}}
\newcommand{\rr}{{\mathrm{r}}}

\newcommand{\rVP}{{\mathrm{VP}}}
\newcommand{\rLE}{{\mathrm{LE}}}
\newcommand{\rLD}{{\mathrm{LD}}}
\newcommand{\rLG}{{\mathrm{LG}}}
\newcommand{\rVE}{{\mathrm{VE}}}
\newcommand{\rVD}{{\mathrm{VD}}}
\newcommand{\rldpc}{{\mathrm{ldpc}}}

\newcommand{\cD}{{\mathcal{D}}}
\newcommand{\bB}{{\mathbb{B}}}
\newcommand{\bC}{{\mathbb{C}}}
\newcommand{\bZ}{{\mathbb{Z}}}
\newcommand{\bR}{{\mathbb{R}}}

%% file: sections/1_abstract.tex
\begin{abstract}
Low Earth orbit (LEO) satellite relays will significantly extend the coverage of mobile networks, enabling users in remote areas to transmit data of real-time events.
Nevertheless, the limited power of user devices and the long distance to satellites lead to low signal-to-noise ratio~(SNR), which results in high error rates and frequent retransmissions, severely hindering the transmissions of high-dimensional data such as videos.
In this paper, we propose a novel method to achieve high error tolerance in satellite-relay video transmissions using generative semantic communications~(GSC).
For the transmitter, we design and optimize a semantic encoder integrating a pre-trained video encoder with a low-density parity-check (LDPC) encoder, efficiently achieving generalizability and enabling forward error correction.
For the receiver, we fine-tune a generative video model using an efficient in-context adaptation algorithm, enabling it to reconstruct videos from error-corrupted semantic information.
Simulation results show that our method achieves 2.5 dB higher video peak SNR than conventional semantic communications at an error rate of 45\%, and remains robust when the error rate exceeds 80\%.
\end{abstract}
 

%% file: sections/2_introduction.tex
\section{Introduction}

Non-terrestrial networks (NTNs) are expected to play a crucial role in future mobile networks~\cite{Azari22CST_Evolution}. 
Within NTNs, the rapid deployment of low Earth orbit (LEO) satellites substantially reduces the latency and path loss in traditional satellite communications. 
By relaying user data through LEO satellites, NTNs can significantly enhance network coverage, enabling users in remote, rural, and maritime areas to share real-time event data, thus providing seamless global connectivity~\cite{Wang22COMM_Ultra}.

Currently, the uplink transmissions from users to LEO satellites still demand large high-gain antennas and high transmit power. 
When user devices, such as mobile phones, directly transmit to LEO satellites, their limited power and the long signal propagation distance result in low signal-to-noise ratios~(SNR) and high error rates~\cite{Xu24TVT_Enhancement}. 
This problem becomes severe for high-dimensional video data, where frequent retransmissions caused by errors lead to low communication efficiency.
Fortunately, the emerging paradigm of semantic communication~(SC)~\cite{Xie21TSP_Deep} can potentially tackle the problem.
In principle, SC uses a neural-model-based semantic encoder to extract low-dimensional semantic information~(SI) from data for transmission~\cite{Hu25COML_Distill}.
By training to handle noisy SI, the semantic encoder can be adapted to low-SNR channels.

Recently, a few studies have leveraged SC to enhance the efficiency of satellite communications, mainly focusing on the downlink, i.e., transmissions from satellites to ground stations.
In~\cite{yin_joint_2025}, a joint source and channel coding~(JSCC) method is proposed for efficient multi-modal data transmission from LEO satellites to ground stations. 
In~\cite{bui_semantic_2025}, a satellite leverages a semantic encoder to compress multi-spectral images at satellites by detecting and encoding only the changed pixels that are vital, thereby saving transmission energy.
In~\cite{chen_free_2025}, a semantic encoder is designed for free-space-optical transmissions of remote sensing images from satellites to ground stations. 
In~\cite{Zheng24JSAC_Semantic}, an SC-assisted framework is designed for satellites to offload user computational tasks to ground stations. 
Moreover, in~\cite{jiang_semantic_2025}, the SC framework leverages a generative model to reconstruct images at the receiver, treating the corrupted SI as conditional information.
Such a generative SC~(GSC) approach seems promising as it leverages the abundant prior knowledge in the generative model for data reconstruction.

However, none of the existing SC works considers the challenging uplink video transmission scenario, where users transmit videos via a satellite relay to a ground station.
In this case, the high dimensionality of video data combined with the severe randomness induced by the high error rate renders it prohibitively costly to train a video semantic encoder with sufficient generalizability. Moreover, adapting a video generative model to be conditioned on corrupted SI also demands substantial data and computational resources.

In this paper, we focus on the uplink video transmission scenario.
Tailored to this scenario, we propose a novel GSC method that achieves high error tolerance.
At the transmitter (Tx), we design and optimize a video semantic encoder that integrates a pre-trained video encoder with an LDPC encoder, ensuring both generalizability and forward error correction capability while significantly simplifying its training.
At the receiver (Rx), we fine-tune a pre-trained video generator using the in-context low-rank adaptation (IC-LoRA) algorithm~\cite{Huang24Arxiv_Context}. 
This efficiently extends the model's spatio-temporal understanding of video features to learn the mapping from error-corrupted SI to original videos.
Our experiments show that the proposed approach achieves notable improvements in reconstruction quality compared with the conventional JSCC method and can preserve perceptual video recognizability even under a high error rate exceeding 80\%.

The rest of this paper is organized as follows.
In Sec.~\ref{sec_sys_mode}, we establish the models for satellite channels and the video GSC system.
In Sec.~\ref{sec_prob_form}, we formulate the optimization problem for the quality of received videos.
To solve the problem, we propose an efficient algorithm in Sec.~\ref{sec_alg_design}, comprising the efficient design of the semantic encoder and the IC-LoRA of the video generator.
Simulation results are provided in Sec.~\ref{sec_eval}, and a conclusion is drawn in Sec.~\ref{sec_conclu}.

%% file: sections/3_systemModel.tex
\section{System Model}\label{sec_sys_mode}
\ifSingleColumn
    \renewcommand{\figwidth}{0.7\linewidth}
\else
    \renewcommand{\figwidth}{0.9\linewidth}
\fi

\begin{figure}[t]
  \centering
  \includegraphics[width=\figwidth]{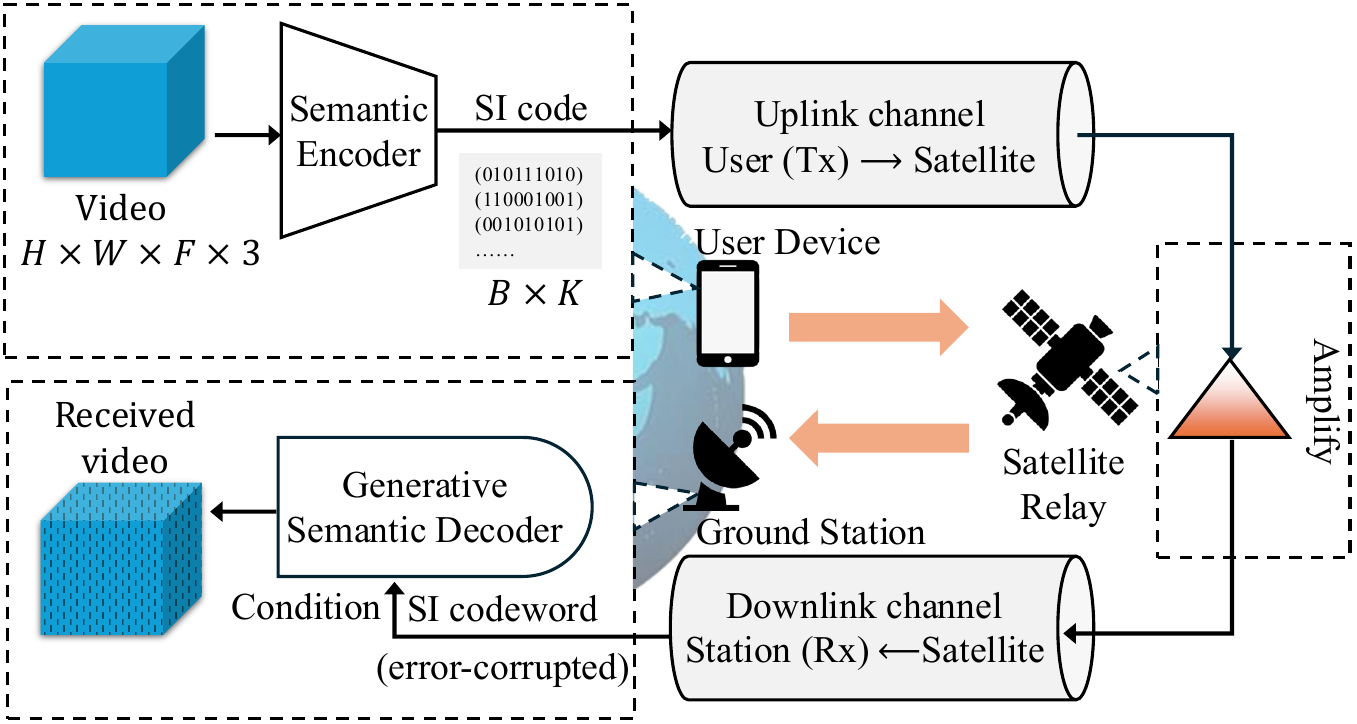}
  \caption{GSC system for video transmission via satellite relay.}
  \label{fig1}
\end{figure}

In this section, we model the GSC system for video uplink via a LEO satellite relay.
As shown in Fig.~\ref{fig1}, the Tx user transmits video data, denoted by $\bm v$, to an Rx ground station via the relay of a LEO satellite.
Without loss of generality, we assume that $\bm v$ comprises $F$ frames, where each frame is an 8-bit RGB image with size $H\times W$, i.e., $\bm v\in\{0,...,255\}^{H\times W\times F \times 3}$.
For transmission efficiency, the user employs a semantic encoder, denoted by $\bm f(\cdot):\{0,...,255\}^{H\times W\times F \times 3}\mapsto \bB^{B\times K}$, encoding a video into $B$ blocks, each block comprising $K$ bits.
The $BK$ bits are collectively referred to as the \emph{SI codeword} of original video $\bm v$.
Each block of bits is mapped to quadrature phase shift keying~(QPSK) symbols, modulated onto a radio frequency~(RF) carrier, and transmitted to the LEO satellite.
The satellite uses the typical \emph{amplify-then-forward} scheme, relaying the SI codeword to the ground station.

Due to the large attenuation and deep fading of the propagation channel to/from the satellite, the received SI codeword suffers from frequent bit errors.
To avoid the long delay caused by retransmissions and the large overhead due to redundant transmissions, the ground station adopts a generative semantic decoder, denoted by $\bm g(\cdot):\bB^{B\times K}\mapsto \{0,...,255\}^{H\times W\times F \times 3}$, reconstructing the original video through a generation process conditioned on the error-corrupted SI codeword.

In the following, we first model the propagation channels and then elaborate on the video transmission in GSC.

%----------------------------------------
\subsection{Channel Model for Satellite Communications}\label{s2ec_chnl_mod}
%----------------------------------------
The propagation channel model comprises two parts: the channel from the user to the satellite, i.e., the \emph{uplink}, and the channel from the satellite to the ground station, i.e., the \emph{downlink}.
For clarity, we focus on the symbol-level equivalent baseband channel model.
In the uplink, denote the transmitted unit-power symbol by $x\in\bC$, $|x|=1$, and the received signal can be expressed as
\begin{equation}
  \label{y_rs}
y_{\rs} = \sqrt{P_{\rt}\cdot {G}_{\rt}\cdot {G}_{\rs} \cdot
  \left({\lambda}/{4\pi d_{\rt\rs}}\right)^{\alpha}}
  \cdot h_{\rt\rs} x + n_{\rs},
\end{equation}
where $P_{\rt}$ is the transmit power of the user, 
${G}_{\rt}$ and ${G}_{\rs}$ are the antenna gains of the user and the satellite, respectively, 
$d_{\rt\rs}$ is the distance between the user and the satellite, 
$\alpha$ is the path-loss exponent, 
$\lambda$ is the RF carrier wavelength, 
$h_{\rt\rs}$ denotes the channel gain, 
and $n_{\rs}$ is the noise at the satellite with zero mean and variance $\sigma_{\rs}^2$.

Based on~\cite{dwivedi2023performance}, we model $h_{\rt\rs}$ as a random variable following the shadowed-Rician distribution, which is a generalized form of the Rician fading model suited for satellite links.
Specifically, denoting $H_{\rt\rs} = \eta_{\rt} | h_{\rt\rs} |^2$ with $\eta_{\rt} = P_{\rt}{G}_{\rt}{G}_{\rs}(\lambda/4\pi)^{\alpha}/\sigma_{\rs}^2$, the probability density function~(PDF) of $H_{\rt\rs}$ can be expressed as
\begin{equation}
  \label{pr_h_rt_rs}
\Pr(H_{\rt\rs}=z) = \varLambda \sum_{\kappa=0}^{m - 1} \frac{\zeta(\kappa)}{\eta_{\rt}^{\kappa + 1}} z^\kappa \exp(- ({\beta - \delta})\cdot z/{\eta_{\rt}} ),
\end{equation}
where $\varLambda = ((2bm)/(2bm + \varOmega))^{m} / 2b$, $\beta = 1 / 2b$, $\delta = {\varOmega}/{(2b)(2bm + \varOmega)}$ and $\zeta(\kappa) = (-1)^\kappa (1 - m)_\kappa \delta^\kappa / (\kappa!)^2$ with $(1 - m)_\kappa$ being the Pochhammer symbol~\cite{Gradshteyn00Tables}. 
Parameters $m$, $b$, and $\varOmega$ characterize the severity of shadowing, whose values are specified in Sec.~\ref{sec_eval}. 

When the satellite receives the uplink signal, it amplifies the signal to its maximum power $P_{\rs}$ and forwards it in the downlink to the Rx ground station. 
Based on~\eqref{y_rs}, the amplification ratio can be calculated as
\beq
\varPsi  = \sqrt{{P_{\rs}}/{\big(\bar{H}_{\rt\rs}d_{\rt\rs}^{-\alpha}\sigma_{\rs}^2+\sigma_{\rs}^2\big)}},
\eeq
where $\bar{H}_{\rt\rs}=\mathbb E[{H}_{\rt\rs}]$ and can be calculated based on~\eqref{pr_h_rt_rs}.
Then, the received signal at the Rx station can be expressed as
\beq
\label{y_rr}
y_{\rr} = \varPsi\cdot\sqrt{ {G} \cdot {G}_{\rr}  \cdot
  \left({\lambda}/{4\pi d_{\rs\rr}}\right)^{\alpha}}
  \cdot h_{\rs\rr} y_{\rs} + n_{\rr},
\eeq
where $d_{\rs\rr}$ is the distance between the satellite and the Rx, and $n_{\rr}$ is the noise at the Rx with zero mean and variance $\sigma_{\rr}^2$.

Based on~\eqref{y_rr}, the SNR for the end-to-end signal transmission from the Tx to the Rx can be calculated by
\begin{equation}\label{equ_snr}
\gamma_{\rt\rr} = \frac{ d_{\rt\rs}^{-\alpha} d_{\rs\rr}^{-\alpha}   H_{\rt\rs} H_{\rs\rr} \varPsi^2 \sigma_{\rs}^2 \sigma_{\rr}^2/P_{\rs}}{d_{\rs\rr}^{-\alpha} H_{\rs\rr} \varPsi^2 \sigma_{\rs}^2 \sigma_{\rr}^2/{P_{\rs}} + \sigma_{\rr}^2},
\end{equation}
where the numerator represents the end-to-end signal power received at the Rx, the first term in the denominator represents the power of the noise forwarded by the satellite, and the second term is the noise power at the Rx.

%----------------------------------------
\subsection{Video Transmission in GSC}\label{s2ec_e2e}
%----------------------------------------
When the Tx user transmits video $\bm v$, it uses the semantic encoder to encode $\bm v$ into SI codeword $\bm c = \bm f(\bm v)\in \bB^{B\cdot K}$. 
Then, the $B$ blocks of bits are sequentially transmitted to the Rx ground station via the relay of the satellite.
Due to the large attenuation and deep fading in the satellite channels, the received SI codeword may suffer from frequent bit errors and non-trivial information loss.
Denoting the lossy channel by function $\bm h: \bB^{B\cdot K}\times \bR \rightarrow \bB^{B\cdot K}$ conditioned on SNR $\gamma_{\rtr}$, the received SI codeword is expressed by $\hat{\bm c}=\bm h(\bm c|\gamma_{\rtr})$.

To tackle the non-trivial information loss, the Rx station uses a generative semantic decoder, denoted by ${\bm g}(\cdot)$, to reconstruct the original video from $\hat{\bm c}$.
Owing to the efficiency and strong performance of diffusion-based conditional video generation~\cite{Xing24ACS_VDM}, we model ${\bm g}(\cdot)$ as a conditional video diffusion model with $\hat{\bm c}$ as the conditioning input.
In this case, ${\bm g}(\cdot)$ is characterized by a conditional noise predictor $\bm\epsilon(\cdot)$ and an iterative denoising procedure.
Based on~\cite{song2020denoising}, from the video at noise level $\tau$, denoted by ${\bm v}_{\tau}$, the denoised video at a smaller noise level $\tau'$ ($\tau',\tau\in[0,1], \tau'<\tau$) is computed as
\beq
\label{equ_denoise}
{{\bm v}_{\tau'}} = \sqrt{\frac{\alpha_{\tau'}}{\alpha_{\tau}}}{{\bm v}_{\tau}} -  \sqrt{\alpha_{\tau'}-\alpha_{\tau}\over \alpha_{\tau}} \cdot \bm\epsilon({\bm v}_{\tau}, \tau|\hat{\bm c}),
\eeq
where $\alpha_{\tau}$ (or $\alpha_{\tau'}$) represents the expected proportion of signal power contributed by noiseless video at level $\tau$ (or $\tau'$).
We note that in~\eqref{equ_denoise}, $\alpha_{\tau}$~($\forall \tau\in[0,1]$) is determined in the training of the diffusion model and commonly treated as a constant.

In practice, the denoising procedure starts from pure noise, i.e., ${\bm v}_1\sim\mathcal N(\bm 0, \bm I)$.
Then, $S$ uniformly spaced levels within $[0,1]$ are selected for iterative denoising.
The final reconstructed video is the denoised video at level $0$, i.e., $\hat{\bm v} =\bm g(\hat{\bm c})={\bm v}_0$.

%% file: sections/4_probFormulation.tex
\section{Problem Formulation}\label{sec_prob_form}

In this section, we formulate the optimization problem for the received video quality in the GSC system.
For the objective function used to evaluate video quality, existing methods use peak SNR~(PSNR) or multi-scale structural similarity index~(MS-SSIM)~\cite{Tung22JSAC_DeepWiVe}. 
Nevertheless, as shown in computer vision studies~\cite{zhang2018unreasonable}, such traditional metrics are generally less effective than the perceptual loss, which is the distance between latent representations extracted by pre-trained neural encoders.
Motivated by this, for the optimization objective, we adopt a \emph{video perceptual loss}~(VPL) associated with a pre-trained video encoder.
Specifically, the video encoder can be represented by $\bm f_{\rVE}: \{0,...,255\}^{H\times W\times F\times 3} \rightarrow \bR^{\frac{H}{\rho_{\rS}}\times \frac{W}{\rho_{\rS}}\times \frac{F}{\rho_{\rT}}\times C}$, where $\rho_{\rS}, \rho_{\rT}\in \mathbb Z$ are the spatial and temporal compression ratios, respectively, and $C$ is the feature dimensionality.
Through the encoding process, each RGB patch of size $\rho_{\rS}\times\rho_{\rS}\times \rho_{\rT}$ in the original video is encoded into $C$ features.

Accordingly, we train the semantic encoder and the generative semantic decoder of the GSC system for the minimization of the VPL, which can be formulated as
\begin{align}
\label{opt_vp}
\min_{\bm f(\cdot), \bm g(\cdot)}&~
{\mathbb E}_{\bm v,\bm h,\gamma_{\rtr}} [L_{\rVP}(\bm v, \hat{\bm v})] \\
&= \mathbb E_{\bm v,\bm h,\gamma_{\rtr}} \big[\| \bm f_{\rVE}(\bm v) - \bm f_{\rVE}\circ\bm g\circ\bm h(\bm f(\bm v)|\gamma_{\rtr})\|_2^2 \big], \nonumber
\end{align}
where $\circ$ is the function composition, $L_{\rVP}(\cdot)$ denotes the VPL, and the expectation is taken over video $\bm v$, channel SNR $\gamma_{\rtr}$, and channel $\bm h(\cdot)$.
We will show in Sec.~\ref{ssec_eval_res} that the VPL is consistent with PSNR and MS-SSIM. 

However, solving~\eqref{opt_vp} faces two challenges. \emph{First}, due to the high dimensionality of video data and the severe randomness caused by the low-SNR channel, training a semantic encoder that can generalize across diverse videos while tolerating potential errors is challenging.
\emph{Second}, training a generative model to handle corrupted SI codewords as conditional information is highly non-trivial, usually requiring training billions of parameters over massive datasets. 
Moreover, the randomness of the corrupted SI codeword further complicates the training process, resulting in prohibitively high resource requirements.

%% file: sections/5_algDesign.tex
\section{Algorithm Design}
\label{sec_alg_design}

In this section, we propose an efficient VPL minimization algorithm for solving~\eqref{opt_vp}.
To avoid the complexity of joint optimization, we decompose~\eqref{opt_vp} into two sub-problems: \emph{semantic encoder optimization} and \emph{generative decoder optimization}.

%========================================
\subsection{Semantic Encoder Optimization}\label{s2ec_seo}
%========================================
We propose an efficient design for the semantic encoder to convert the challenging optimization into a more tractable problem.
Specifically, we incorporate the pre-trained video encoder $\bm f_{\rVE}(\cdot)$ as part of the semantic encoder.
By this means, we naturally ensure high generalizability across diverse videos.
In addition, the LDPC encoder is integrated into the semantic encoder, which not only provides strong forward error correction (FEC) capability but also ensures compatibility with prevalent communication infrastructures.

We use a latent encoder to bridge the video encoder and the LDPC encoder, and design the semantic encoder as
\beq\label{equ_f}
\bm f(\bm v) = \bm f_{\rldpc} \circ \bm f_{\rLE}\circ\bm f_{\rVE}(\bm v),
\eeq
where $\bm f_{\rLE}:\bR^{\frac{H}{\rho_{\rS}}\times \frac{W}{\rho_{\rS}}\times \frac{F}{\rho_{\rT}}\times C}\rightarrow \{1,...,Q\}^M$ denotes the \emph{latent encoder}, with quantization level $Q$ and SI vector length $M$, mapping a latent tensor $\bm u$ to a discrete SI vector $\bm s = \bm f_{\rLE}(\bm u)$, 
and $\bm f_{\rldpc}: \{1,...,Q\}^M \rightarrow \bB^{BK}$ denotes the LDPC encoder.

For the latent encoder, we design it as a neural model with typical 3D convolutional layers and trainable parameter matrices, which we collectively denote by a shorthand notation $\{\bm W_{\rLE}\}$.
As for the architectural hyper-parameters of the latent encoder, we focus on $Q$ and $M$ due to their dominant roles in the information capacity of the SI vector.
For the LDPC encoder, it is treated as a deterministic function.
Specifically, the LDPC encoder takes an SI vector, evenly divides its $M$ elements into $B$ groups, and encodes each group into a block of $K$ bits with code rate $R$.
Within each block, $RK$ bits are information bits, and $(1-R)K$ bits are parity bits.

When the Rx station receives the SI codeword, it leverages an LDPC decoder $\bm f_{\rldpc}^{-1}(\cdot)$ to correct bit errors and reconstruct the SI vector.
The probability that a block remains erroneous after LDPC decoding is a function of $K$, $R$, and $\gamma_{\rtr}$, denoted by $p(K, R,\gamma_{\rtr})$, which is referred to as the block error rate~(BLER) and can be estimated through Monte Carlo simulation.
If a block in the SI codeword is erroneous, all the SI vector elements contained in the block are lost, resulting in the reconstructed SI vector being corrupted.
We represent the relation of the $M$ elements of the SI vector included in the $n$-th block of the SI codeword by a mask vector $\bm m_n \in \mathbb{B}^M$.
Then, the corrupted SI vector can be expressed as
\beq
\hat{\bm s} = \sum_{n=1}^N (1-X_n)\cdot \bm s\odot \bm m_n,
\eeq
where $\odot$ denotes the element-wise product, and $X_n\in\bB$ is Bernoulli random variable with $\Pr(X_n=1)=p(K, R,\gamma_{\rtr})$.

To facilitate optimization of the latent encoder, we design a corresponding latent decoder $\bm f_{\rLD}:\{1,...,Q\}^M\rightarrow \bR^{\frac{H}{\rho_{\rS}}\times \frac{W}{\rho_{\rS}}\times \frac{F}{\rho_{\rT}}\times C}$ with trainable parameter matrices $\{\bm W_{\rLD}\}$, which has a similar architecture to $\bm f_{\rLE}(\cdot)$.
The latent decoder reconstructs a latent tensor from the corrupted SI vector, i.e., $\hat{\bm u}=\bm f_{\rLD}(\hat{\bm s})$.
We refer to $\hat{\bm u}$ as the \emph{corrupted latent tensor}.

In accordance with~\eqref{opt_vp}, we minimize the distance between $\bm u$ and $\hat{\bm u}$.
As the video encoder and the LDPC codec are fixed, the challenging problem~\eqref{opt_vp} is simplified to optimizing the hyper-parameters $Q$ and $M$ and the trainable parameter matrices $\{\bm W_{\rLE},\bm W_{\rLD}\}$, which can be formulated as
\begin{subequations}\label{equ_opt_z_full}
\begin{align}\label{equ_opt_z}
\min_{Q,M\in \bZ}&~J(Q,M)= \min_{\{\bm W_{\rLE}, \bm W_{\rLD}\}}\!\sum_{\bm u\in\cD_{\rL}}{\mathbb E}[\|\hat{\bm u} - \bm u\|_2^2], \\
\label{equ_opt_z_2}
& \log_2(Q) \cdot M\leq B\cdot K\cdot R,
\end{align}
\end{subequations}
where $\cD_{\rL}$ denotes a dataset of latent tensors for $D$ videos encoded by $\bm f_{\rVE}(\cdot)$, and the expectation in~\eqref{equ_opt_z} is taken over the random block errors and channel SNR.
The inner problem \eqref{equ_opt_z} can be solved by standard neural model optimizers such as Adam~\cite{Kingma14Arxiv_Adam}.
Constraint~\eqref{equ_opt_z_2} limits the total number of output information bits of the latent encoder.

To solve~\eqref{equ_opt_z_full}, we find that, for a fixed $Q$, increasing $M$ improves $J(Q,M)$ as more information can be retained after latent encoding.
Therefore, the optimal $M$ can be obtained by
\beq\label{equ_m_star}
M^*(Q) = \arg\max_{M'\in\bZ} M', \text{ s.t. } M'\leq B\cdot K\cdot R / \log_2{Q},
\eeq
and the optimal $Q$ can be obtained by
\beq\label{equ_q_star}
Q^* = \arg\min_{Q\in \bZ} J(Q,M^*(Q)),
\eeq
which can be solved efficiently by enumerating a few small integers~(e.g., integers less than 5), as shown in Sec.~\ref{ssec_eval_res}.

\subsection{Generative Decoder Optimization}
To avoid the prohibitive cost of training a generative model from scratch, the most intuitive idea is to adopt a pre-trained video generator and adapt it to handle corrupted SI as conditioning input.
However, adding conditions to a video generator generally requires complex architectural modifications and a large amount of training, which is still difficult in practice.
Moreover, the fact that the reconstruction quality in~\eqref{opt_vp} is measured in the latent space of $\bm f_{\rVE}(\cdot)$ further complicates the optimization of the video generator.

To handle this difficulty, we directly adopt a noise predictor $\bm\epsilon_{\rLG}(\cdot)$ pre-trained on latent tensors encoded by $f_{\rVE}(\cdot)$ and leverage the IC-LoRA algorithm~\cite{Huang24Arxiv_Context} to enable $\bm\epsilon_{\rLG}(\cdot)$ to be conditioned on corrupted latent tensor $\hat{\bm u}$.
The key idea can be explained as follows.
Through comprehensive pre-training, the noise predictor inherently possesses the ability to capture spatio-temporal relationships among the features in a latent tensor, which can be efficiently adapted to the mapping from corrupted latent tensors to original ones.

Accordingly, we fine-tune $\bm\epsilon_{\rLG}(\cdot)$ to generate the original latent tensor $\bm u$ given its corrupted version $\hat{\bm u}$, which is obtained by $\hat{\bm u}=\bm f^{-1}_{\rldpc}\circ\bm f_{\rLD}(\hat{\bm c})$.
Essentially, this converts the objective in~\eqref{opt_vp} to the minimization of the noise prediction loss for noisy $\bm u$.
Inspired by the IC-LoRA algorithm, we directly use $\bm\epsilon_{\rLG}(\cdot)$ to handle the concatenation of noisy $\bm u$ and $\hat{\bm u}$ and calculate the noise prediction loss over $\bm u$, which can be expressed as 
\beq
L(\bm u, \hat{\bm u};\{\bm W_{\rLG}\}) = {\mathbb E}_{\bm\varepsilon,\tau}[ \| \bm \varepsilon - \bm \varGamma(\bm\epsilon_{\rLG}({\bm u}_{\tau}\oplus\hat{\bm u}_{\tau},\tau) ) \|_2^2 ],
\eeq
where $\{\bm W_{\rLG}\}$ denotes the pre-trained parameter matrices of $\bm\epsilon_{\rLG}(\cdot)$,
$\oplus$ denotes the concatenation along spatial dimensions,
$\bm\varGamma(\cdot)$ is the cut-off operation for the latter half along spatial dimensions,
and ${\bm u}_\tau$ and $\hat{\bm u}_\tau$ are calculated by
\beq
{\bm u}_{\tau} = \sqrt{\alpha_{\tau}}{\bm u} + \sqrt{1-\alpha_{\tau}} {\bm\varepsilon},~\hat{\bm u}_{\tau} = \sqrt{\alpha_{\tau}}\hat{\bm u} + \sqrt{1-\alpha_{\tau}} \hat{\bm\varepsilon},
\eeq
where $\hat{\bm\varepsilon}$ and ${\bm\varepsilon}$ are zero-mean unit-variance Gaussian noise.

Furthermore, to avoid the complexity in fine-tuning the complete parameter matrices of $\bm\epsilon_{\rLG}(\cdot)$, we only train LoRA matrices added to $\{\bm W_{\rLG}\}$.
In particular, for each pre-trained parameter matrix $\bm W_{\rLG}\in\bR^{O\times I}$ with $O$ and $I$ denoting the output and input dimensions, its LoRA matrices are $\bm B\in\bR^{O\times r}$ and $\bm A\in\bR^{r\times I}$, where $r$ represents the rank value ($r\ll O,I$).
Based on~\cite{dekap_g}, training LoRA matrices has been shown to effectively inject new knowledge into pre-trained generative neural models.
Therefore, the training can be expressed as 
\beq\label{equ_lora_ft}
\min_{\{\bm B, \bm A\}}~\sum_{\bm u\in\cD'_{\rL}} L(\bm u, \hat{\bm u}; \{\bm W_{\rLG} + \bm B\bm A\}),
\eeq
where $\cD'_{\rL}$ is a dataset of latent tensors different from $\cD_{\rL}$.

Problem~\eqref{equ_lora_ft} can be solved by standard neural model optimizers such as Adam~\cite{Kingma14Arxiv_Adam}.
At deployment, when receiving $\hat{\bm c}$, the Rx first obtains $\hat{\bm u} = \bm f_{\rLD}\circ\bm f^{-1}_{\rldpc}(\hat{\bm c})$, and then uses $\bm\epsilon_{\rLG}(\cdot)$ to generate a reconstructed latent tensor $\tilde{\bm u}$, following a similar denoising procedure to~\eqref{equ_denoise}.
Then, it decodes $\tilde{\bm u}$ and obtains the reconstructed video $\hat{\bm v} = \bm f_{\rVD}(\tilde{\bm u})$.
In summary, the efficient VPL minimization algorithm is outlined in Algorithm~\ref{alg_vpl}.
\input{sections/alg_1.tex}

%% file: sections/alg_1.tex
\ifSingleColumn
\begin{figure}[H]
\else
\begin{figure}[H]
\fi
\begin{algorithm}[H]
\ifSingleColumn
\normalsize
\else
\small
\fi
\caption{Efficient VPL minimization algorithm.}
\label{alg_vpl}
\begin{algorithmic} [1]
\Require Latent tensor datasets $\cD_{\rL}$ and $\cD'_{\rL}$; pre-trained video encoder and decoder $\bm f_{\rVE}(\cdot)$, $\bm f_{\rVD}(\cdot)$; pre-trained noise predictor $\bm\epsilon_{\rLG}(\cdot)$; rank value for LoRA matrices $r$; SI codeword parameters $B$, $K$, $R$; coefficients in the satellite channel model $m$, $b$, $\varOmega$.
\Statex {\#~\textbf{\textit{--- Semantic Encoder Optimization ---}}\#}
\State Solve~\eqref{equ_m_star} and~\eqref{equ_q_star} and obtain the latent encoder and decoder with optimized parameters, i.e., $\bm f^*_{\rLE}(\cdot)$ and $\bm f^*_{\rLD}(\cdot)$.
\State Obtain optimized semantic encoder $\bm f^* =\bm f_{\rldpc}\circ\bm f^*_{\rLE}\circ \bm f_{\rVE}$.
\Statex {\#~\textbf{\textit{--- Generative Decoder Optimization ---}}\#}
\State With $\bm f^*_{\rLE}(\cdot)$ and $\bm f^*_{\rLD}(\cdot)$, obtain corrupted latent tensor $\hat{\bm u}$ for each $\bm u\in\cD'_{\rL}$ by $\hat{\bm u} = \bm f^*_{\rLD}\circ \bm h\circ \bm f^*_{\rLE}(\bm u)$.
\State Optimize LoRA matrices for the pre-trained parameters of $\bm\epsilon_{\rLG}(\cdot)$ by solving~\eqref{equ_lora_ft}, and obtain $\bm \epsilon_{\rLG}^*(\cdot)$.
\State Obtain optimized generative semantic decoder $\bm g^*(\cdot)$, which comprises $\bm \epsilon_{\rLG}^*(\cdot)$, $\bm f^*_{\rLD}\circ\bm f_{\rldpc}^{-1}(\cdot)$, and $\bm f_{\rVD}(\cdot)$.
\end{algorithmic}
\end{algorithm}
\end{figure}

%% file: sections/6_evaluation.tex
\section{Evaluation}\label{sec_eval}

In this section, we present the experimental setup and the evaluation results.
\subsection{Experimental Setup}

We simulate a LEO satellite channel working in the Ku-band centered at $10.7$~GHz using typical setup parameters following the 3GPP standard~\cite{3GPP_TR_38_811_V15_4_0}.
The distances from the satellite to the user and to the ground station are assumed to be the same, i.e., $d_{\rt\rs}=d_{\rs\rr}=d$, with $d=600$~\!km by default.
As for LDPC, we adopt the DVB-S2 standard~\cite{ETSI_EN_302_307_1} with a short block length of $K=16{,}200$~\!bits and a low code rate of $R=1/2$, and use QPSK for digital modulation.

\ifSingleColumn
    \renewcommand{\figwidth}{0.7\linewidth}
\else
    \renewcommand{\figwidth}{0.95\linewidth}
\fi
\begin{figure}[t] 
    \centering
    \includegraphics[width=\figwidth]{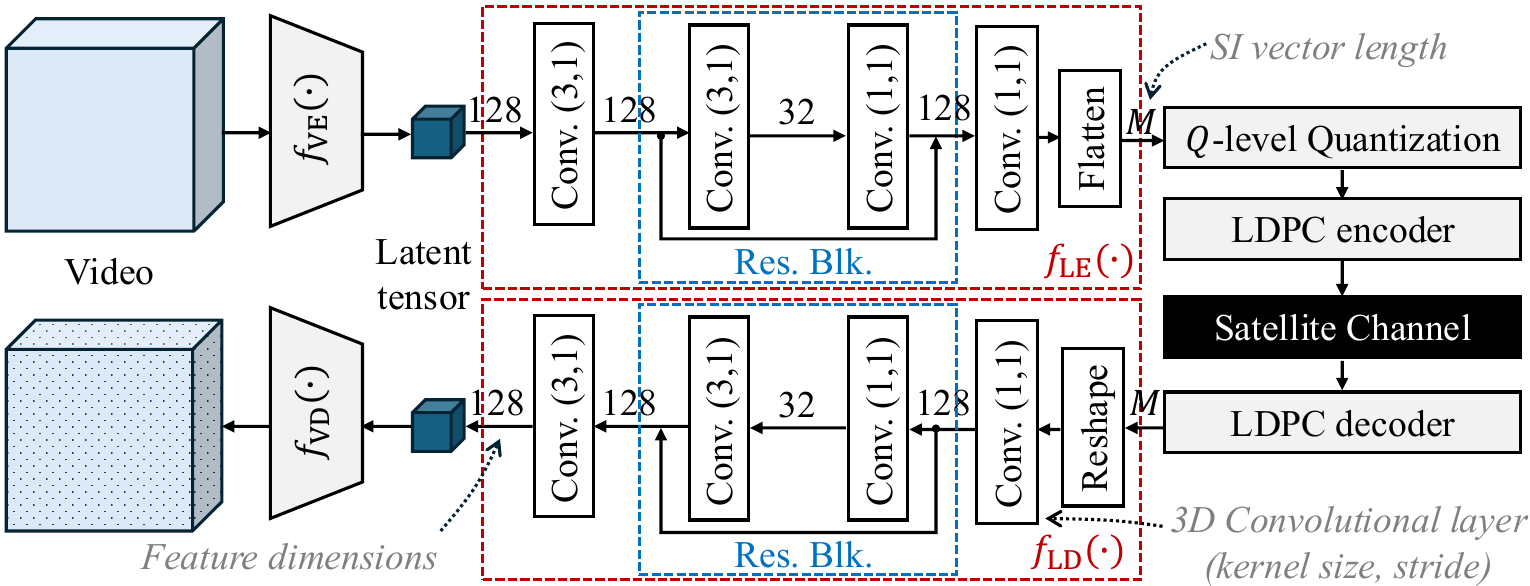}
    \caption{Neural model architecture of the latent encoder and decoder.}
    \label{fig_net_arch}
\end{figure}

\input{sections/table_1}

Each video handled by the GSC system comprises $49$ frames. Each frame has a spatial resolution of $512 \times 512$ pixels and 8-bit RGB channels. 
With $\bm f(\cdot)$, each video is mapped into $B=85$ blocks of bits, resulting in a compression rate of $1.3\%$.
More specifically, for $\bm f_{\rVE}(\cdot)$ and $\bm f_{\rVD}(\cdot)$, we adopt the video codec proposed in~\cite{Hacohen24Arxiv_LTX} for its high efficiency, which maps a video into $16\times 16\times 7$ latent feature vectors.
Since the video codec has already extracted deeply processed features, we adopt a simple and highly efficient architecture for the latent codec as illustrated in Fig.~\ref{fig_net_arch}.
The considered quantization levels include $Q\in\{2,3,4,5\}$, whereas $Q>5$ is unnecessary as shown later in Fig.~\ref{fig_exp3_b}.

For training the semantic encoder, we sample $1{,}000$ videos from the PE video dataset~\cite{bolya2025perception} and encode them with $\bm f_{\rVE}(\cdot)$ to construct $\cD_{\rL}$. Based on $\cD_{\rL}$, the latent codec is trained by~\eqref{equ_opt_z_full}, yielding $Q^*=3$ and $M^*=433{,}664$ for the default $d$. Another $200$ videos are then sampled and encoded to form $\cD_{\rL}'$, which is used to train the generative semantic decoder via~\eqref{equ_lora_ft} with the LoRA rank $r=128$. 
Finally, $30$ videos not included in $\cD_{\rL}$ or $\cD'_{\rL}$ are sampled to evaluate the reconstruction quality.
The detailed parameters are summarized in Tab.~\ref{tab_1}.

\subsection{Evaluation Results}\label{ssec_eval_res}

\ifSingleColumn
    \renewcommand{\figwidth}{0.25\linewidth}
\else
    \renewcommand{\figwidth}{0.32\linewidth}
\fi

\begin{figure}[!t]
    \centering 
    \begin{subfigure}{\figwidth}
        \centering
        \includegraphics[width=\linewidth]{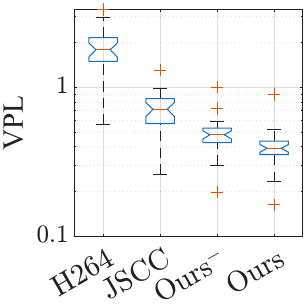}  
        \caption{}
        \label{fig_exp1_a}
    \end{subfigure}
    \ifSingleColumn
    \hspace{3ex}
    \else
    \hspace{-1ex}
    \fi
    \begin{subfigure}{\figwidth}
        \centering
        \includegraphics[width=\linewidth]{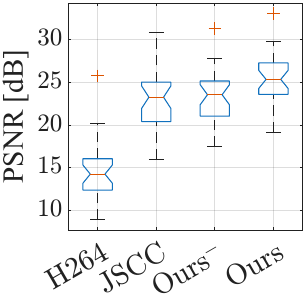}  
        \caption{}
        \label{fig_exp1_b}
    \end{subfigure}
    \ifSingleColumn
    \hspace{3ex}
    \else
    \hspace{-1ex}
    \fi
    \begin{subfigure}{\figwidth}
        \centering
        \includegraphics[width=\linewidth]{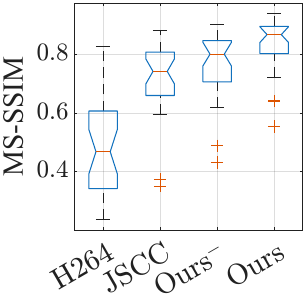}  
        \caption{}
        \label{fig_exp1_c}
    \end{subfigure}
    \caption{Reconstruction quality comparison in terms of (a) VPL, (b) PSNR, and (c) MS-SSIM.}
    \label{fig_exp1}
\end{figure} 

\ifSingleColumn
    \renewcommand{\figwidth}{0.7\linewidth}
\else
    \renewcommand{\figwidth}{0.99\linewidth}
\fi
\begin{figure}[t] 
    \centering
    \includegraphics[width=\figwidth]{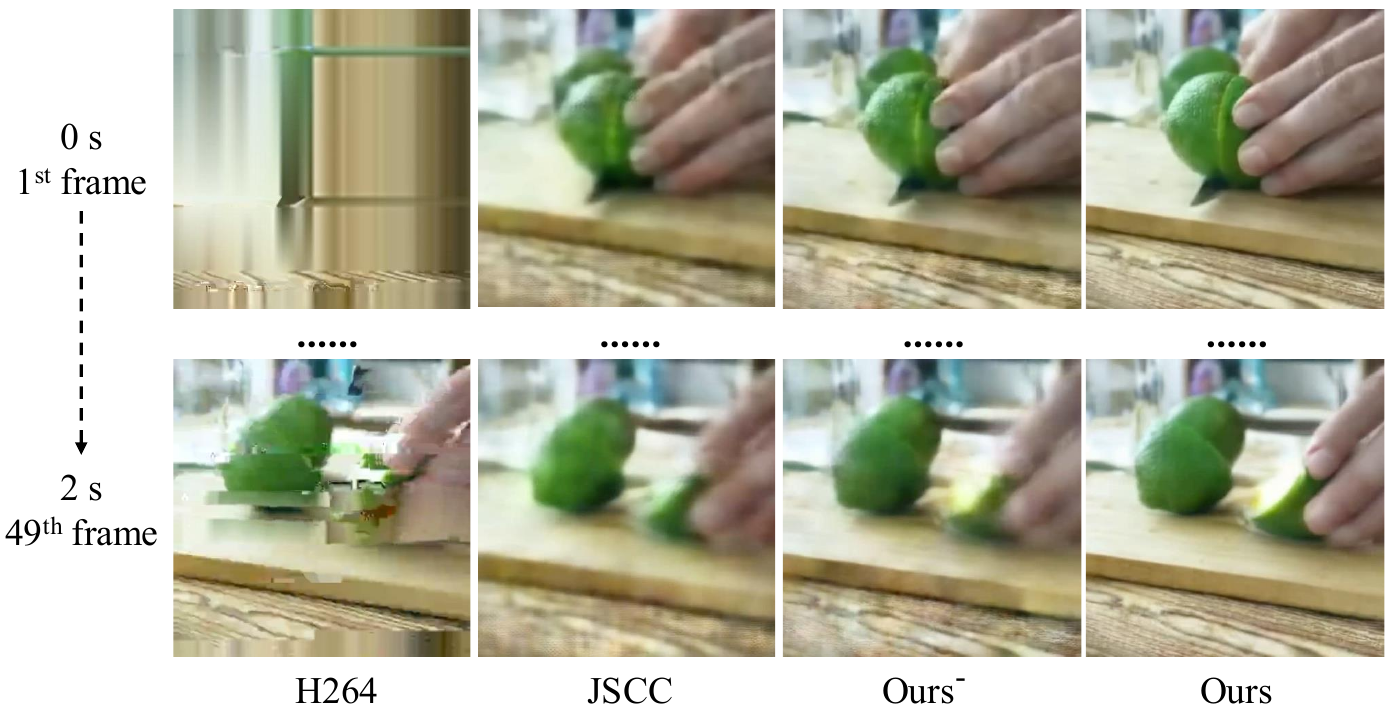}
    \caption{Visual comparison between reconstructed videos.}
    \label{fig_exp2}
\end{figure}

\ifSingleColumn
    \renewcommand{\figwidth}{0.33\linewidth}
\else
    \renewcommand{\figwidth}{0.5\linewidth}
\fi

We compare our method with the conventional JSCC methods in~\cite{yin_joint_2025} and~\cite{Tung22JSAC_DeepWiVe}, as well as with the method based on the prevalent H.264 standard~\cite{H264}.
For fairness, the JSCC method also employs the pre-trained video codec and a latent codec similar to Fig.~\ref{fig_net_arch}.
However, in the JSCC method, latent tensors are directly mapped to QPSK symbols without LDPC coding. 
In the H.264 method, videos are directly encoded by the H.264 standard and transmitted through the same LDPC codec as in our method. 
No generative semantic decoder is employed in the JSCC and H.264 methods.
The bit error rate for H.264 is set to $10^{-3}$, as larger error rates lead to completely undecodable videos.
Moreover, to validate the performance gains of our LDPC-integrated semantic encoder compared to JSCC, we compare a variant of our method without generation process, denoted as \emph{Ours$^{-}$}.
Our complete method is denoted as \emph{Ours}.

Figs.~\ref{fig_exp1_a},~\ref{fig_exp1_b}, and~\ref{fig_exp1_c} compare the reconstructed video quality between our method and the others in terms of the VPL, PSNR, and the MS-SSIM, respectively, and the reconstructed frames are shown in Fig.~\ref{fig_exp2}.
The consistency between the adopted VPL and PSNR and MS-SSIM metrics justifies the objective function in~\eqref{opt_vp}. 
More importantly, our method leads to the lowest VPL and highest PSNR and MS-SSIM values compared to the others.
Compared to \emph{JSCC}, our system improves PSNR and MS-SSIM by an average of $2.5$~\!dB and $19\%$, respectively.
Moreover, \emph{Our$^-$} outperforming \emph{JSCC} validates the benefit of incorporating the LDPC in the semantic encoder.
Furthermore, all the SC-based methods outperform \emph{H.264} even though the error rate in \emph{H.264} is substantially lower.
This demonstrates the high error tolerance of SC-based methods.

\ifSingleColumn
    \renewcommand{\figwidth}{0.4\linewidth}
\else
    \renewcommand{\figwidth}{0.5\linewidth}
\fi
\begin{figure}[!t]
    \centering 
    \begin{subfigure}{\figwidth}
        \centering
        \includegraphics[width=\linewidth]{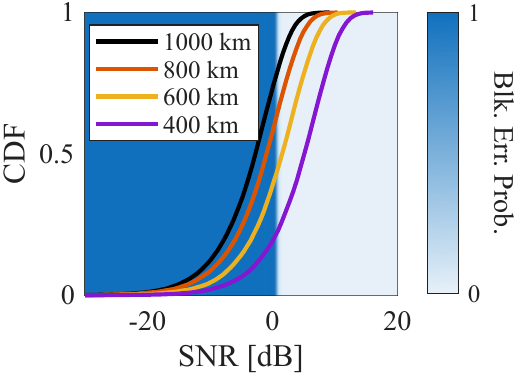}  
        \caption{}
        \label{fig_exp3_a}
    \end{subfigure}
    \ifSingleColumn
    \hspace{3ex}
    \else
    \hspace{-2ex}
    \fi
    \begin{subfigure}{\figwidth}
        \centering
        \includegraphics[width=0.9\linewidth]{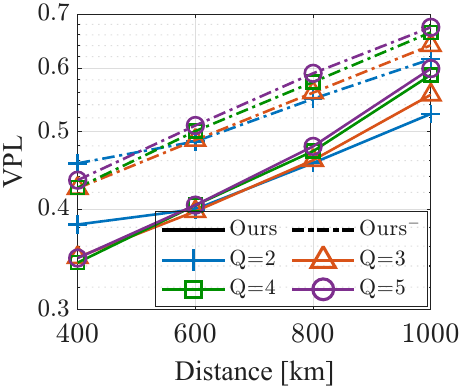}  
        \vspace{-.5ex}
        \caption{}
        \label{fig_exp3_b}
    \end{subfigure}
    \caption{(a) CDF of the channel SNR given different $d$. (b) VPL of the received videos versus $d$, given different $Q$.}
    \label{fig_exp3}
\end{figure} 

\ifSingleColumn
    \renewcommand{\figwidth}{0.8\linewidth}
\else
    \renewcommand{\figwidth}{1\linewidth}
\fi
\begin{figure}[t] 
    \centering
    \includegraphics[width=\figwidth]{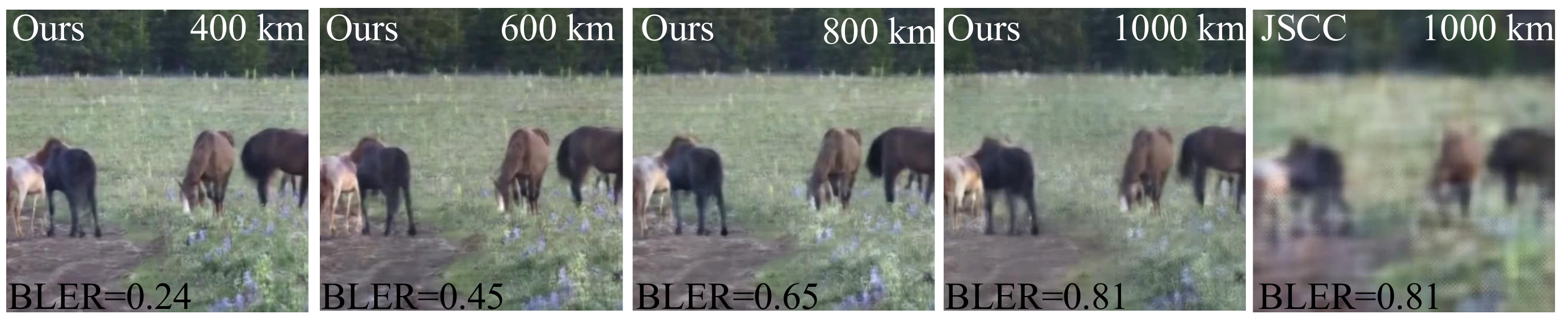}
    \caption{Reconstructed frames at different distance $d$.}
    \label{fig_exp4}
\end{figure}

In Figs.~\ref{fig_exp3_a} and~\ref{fig_exp3_b}, we take a closer look at the impact of distance $d$ and the quantization level $Q$.
Fig.~\ref{fig_exp3_a} illustrates the cumulative distribution functions~(CDF) of SNR $\gamma_{\rtr}$ given different $d$.
The color map illustrates the BLER of the LDPC codec.
The BLER changes rapidly from 1 to 0 at around $1$~\!dB SNR, known as the \emph{cliff effect}.
As $d$ increases from $400$ to $1{,}000$~\!km, the BLER grows rapidly from $24\%$ to $81\%$.
Fig.~\ref{fig_exp3_b} shows that $Q$ has a prominent impact on the resulting VPL, especially when $d=1{,}000$~\!km.
It can be observed that enumerating over $Q\in\{2,3,4,5\}$ is sufficient for solving~\eqref{equ_q_star} when $d\in[400,1000]$~\!km.
The optimal $Q$ varies with $d$:
When $d=400$~\!km, $Q*=4$, whereas when $d=1{,}000$~\!km, $Q*=2$.
This is probably because when BLER is large, it is more important to have more correctly received elements of the SI vector; when BLER is small, it is more important for each element of the SI vector to have higher expressiveness.

Finally, Fig.~\ref{fig_exp4} shows the middle frame of the reconstructed videos obtained by our method and \emph{JSCC} at different $d$.
It can be observed that even at $d=1{,}000$~\!km with a BLER of $81\%$, the video reconstructed by our method is still highly recognizable and has higher quality than that of \emph{JSCC}, demonstrating the higher error tolerance of our method.

%% file: sections/table_1.tex
\ifSingleColumn
    \renewcommand{\figwidth}{0.65\columnwidth}
\else
    \renewcommand{\figwidth}{\columnwidth}
\fi
\begin{table}[]
\caption{Simulation Parameters}
\label{tab_1}
\centering
\resizebox{\figwidth}{!}{%
\begin{tabular}{|l|l|l|l|l|l|}
\hline
\textbf{Parameter} & \textbf{Value} & \textbf{Parameter}      & \textbf{Value} & \textbf{Parameter} & \textbf{Value} \\ \hline
\textbf{$\lambda$} & 2.8 cm         & $d_{\rt\rs},d_{\rs\rr}$ & 600 km         & $G_{\rt}$          & 1 W            \\ \hline
$G_{\rs}$                       & 30 W    & $G_{\rt}$   & 3 dBi & $G_{\rs}, G_{\rr}$ & 50 dBi \\ \hline
$\sigma_{\rs}^2,\sigma_{\rr}^2$ & -98 dBm & $m$         & 2     & $b$                & 0.063  \\ \hline
$\varOmega$                     & 0.0005  & $S$ & 50    & $K$                & 16,200 \\ \hline
$R$                             & 1/2     & $H$,$W$     & 512   & $F$                & 49     \\ \hline
$C$                             & 128     & $B$         & 85    & $\rho_{\rS}$       & 32     \\ \hline
$\rho_{\rT}$                    & 7       & $r$         & 128   &    $D$                &    1{,}000    \\ \hline
\end{tabular}%
}
\end{table}

%% file: sections/7_conclusion.tex
\section{Conclusion}\label{sec_conclu}

In this paper, we have designed a GSC system for user uplink video transmission with a satellite relay, which achieves high error tolerance.
We have proposed an algorithm to efficiently optimize the semantic encoder and the generative semantic decoder for VPL minimization.
Evaluation results validate that our proposed method outperforms the conventional JSCC method by 2.5~\!dB PSNR and 19\% MS-SSIM, respectively, and maintains high perceptual recognizability of received videos even under a high BLER of 81\%.

%% file: bibilio.bib
@article{dekap_g,
  title={{Distillation-Enabled Knowledge Alignment for Generative Semantic Communications of AIGC Images}},
  author={Hu, Jingzhi and Li, Geoffrey Ye},
  journal={arXiv:2506.19893},
  month=jan,
  year={2026}
}

@ARTICLE{H264,
  author={Wiegand, T. and Sullivan, G.J. and Bjontegaard, G. and Luthra, A.},
  journal={IEEE Trans. Circuits Syst. Video Technol.}, 
  title={{Overview of the H.264/AVC video coding standard}}, 
  year={2003},
  volume={13},
  number={7},
  pages={560-576},
  month=jul,}

@article{Hacohen24Arxiv_LTX,
  title={{Ltx-video: Realtime video latent diffusion}},
  author={HaCohen, Yoav and Chiprut, Nisan and Brazowski, Benny and Shalem, Daniel and Moshe, Dudu and Richardson, Eitan and Levin, Eran and Shiran, Guy and Zabari, Nir and Gordon, Ori and others},
  journal={arXiv:2501.00103},
  month=dec,
  year={2024}
}

@article{bolya2025perception,
  title={{Perception encoder: The best visual embeddings are not at the output of the network}},
  author={Bolya, Daniel and Huang, Po-Yao and Sun, Peize and Cho, Jang Hyun and Madotto, Andrea and Wei, Chen and Ma, Tengyu and Zhi, Jiale and Rajasegaran, Jathushan and Rasheed, Hanoona and others},
  journal={arXiv:2504.13181},
  month=apr,
  year={2025}
}

@standard{ETSI_EN_302_307_1,
  author       = {{ETSI}},
  title        = {{Digital Video Broadcasting (DVB-S2) Standard}},
  number       = {EN 302 307-1},
  organization = {European Telecommunications Standards Institute},
  year         = {2014}
}

@article{Kingma14Arxiv_Adam,
  title={{Adam: A method for stochastic optimization}},
  author={Kingma, Diederik P and Ba, Jimmy},
  journal={arXiv:1412.6980},
  month=dec,
  year={2014}
}

@techreport{3GPP_TR_38_811_V15_4_0,
  author  = {3GPP},
  title        = {{Study on New Radio (NR) to support non-terrestrial networks}},
  type         = {Technical Report, TR\,38.811},
  year         = {2020},
  month        = sep,
  note         = {{Release 15}},
}

@ARTICLE{Tung22JSAC_DeepWiVe,
  author={Tung, Tze-Yang and Gündüz, Deniz},
  journal={IEEE J. Sel. Areas Commun.}, 
  title={{DeepWiVe: Deep-Learning-Aided Wireless Video Transmission}}, 
  year={2022},
  month=sep,
  volume={40},
  number={9},
  pages={2570-2583},
}

@article{Xing24ACS_VDM,
author = {Xing, Zhen and Feng, Qijun and Chen, Haoran and Dai, Qi and Hu, Han and Xu, Hang and Wu, Zuxuan and Jiang, Yu-Gang},
title = {A Survey on Video Diffusion Models},
year = {2024},
volume = {57},
number = {2},
journal = {ACM Comput. Surv.},
month = nov,
pages = {1--42},
}

@book{Gradshteyn00Tables,
  author    = {I. S. Gradshteyn and I. M. Ryzhik},
  title     = {Tables of Integrals, Series, and Products},
  publisher = {Academic Press},
  address   = {New York, NY, USA},
  year      = {2000},
}

@article{Huang24Arxiv_Context,
  title={{In-context LoRA for diffusion transformers}},
  author={Huang, Lianghua and Wang, Wei and Wu, Zhi-Fan and Shi, Yupeng and Dou, Huanzhang and Liang, Chen and Feng, Yutong and Liu, Yu and Zhou, Jingren},
  journal={arXiv:2410.23775},
  month=nov,
  year={2024}
}

@ARTICLE{Hu25COML_Distill,
  author={Hu, Jingzhi and Li, Geoffrey Ye},
  journal={IEEE Commun. Lett.}, 
  title={{Distillation-Enabled Knowledge Alignment Protocol for Semantic Communication in AI Agent Networks}}, 
  month=aug,
  year={2025},
  volume={29},
  number={11},
  pages={2541-2545},
    month=nov,
}

@ARTICLE{Xie21TSP_Deep,
  author={Xie, Huiqiang and Qin, Zhijin and Li, Geoffrey Ye and Juang, Biing-Hwang},
  journal={IEEE Trans. Signal Process.}, 
  title={Deep Learning Enabled Semantic Communication Systems}, 
  month=apr,
  year={2021},
  volume={69},
  pages={2663-2675},}

@ARTICLE{Xu24TVT_Enhancement,
  author={Xu, Zhuoao and Chen, Gaojie and Fernandez, Ryan and Gao, Yue and Tafazolli, Rahim},
  journal={IEEE Trans. Veh. Technol.}, 
  title={Enhancement of Direct LEO Satellite-to-Smartphone Communications by Distributed Beamforming}, 
  year={2024},
  volume={73},
  number={8},
  pages={11543-11555},
  month=aug}

@ARTICLE{Wang22COMM_Ultra,
  author={Wang, Ruibo and Kishk, Mustafa A. and Alouini, Mohamed-Slim},
  journal={IEEE Communications Magazine}, 
  title={Ultra-Dense LEO Satellite-Based Communication Systems: A Novel Modeling Technique}, 
  year={2022},
  volume={60},
  number={4},
  pages={25-31},
  month=apr}

@ARTICLE{Azari22CST_Evolution,
  author={Azari, M. Mahdi and Solanki, Sourabh and Chatzinotas, Symeon and Kodheli, Oltjon and Sallouha, Hazem and Colpaert, Achiel and Mendoza Montoya, Jesus Fabian and Pollin, Sofie and Haqiqatnejad, Alireza and Mostaani, Arsham and Lagunas, Eva and Ottersten, Bjorn},
  journal={IEEE Commun. Surv. Tutor.}, 
  title={{Evolution of Non-Terrestrial Networks From 5G to 6G: A Survey}}, 
  year={2022},
  volume={24},
  number={4},
  pages={2633-2672},
  month={Fourthquarter},
}

@ARTICLE{Zheng24JSAC_Semantic,
  author={Zheng, Guhan and Ni, Qiang and Navaie, Keivan and Pervaiz, Haris},
  journal={IEEE J. Sel. Areas Commun.}, 
  title={Semantic Communication in Satellite-Borne Edge Cloud Network for Computation Offloading}, 
  month=may,
  year={2024},
  volume={42},
  number={5},
  pages={1145-1158},
  }

@inproceedings{zhang2018unreasonable,
  title = {The Unreasonable Effectiveness of Deep Features as a Perceptual Metric},
  author = {Zhang, Richard and Isola, Phillip and Efros, Alexei A. and Shechtman, Eli and Wang, Oliver},
  booktitle = {Proc. IEEE/CVF CVPR},
  address={Salt Lake City, UT, USA},
  month={Jun},
  year = {2018}
}

@article{song2020denoising,
  title = {Denoising Diffusion Implicit Models},
  author = {Song, Jiaming and Meng, Chenlin and Ermon, Stefano},
  journal = {arXiv preprint arXiv:2010.02502},
  year = {2020}
}

@article{dwivedi2023performance,
  title={Performance analysis of LEO satellite-based IoT networks in the presence of interference},
  author={Dwivedi, Ayush Kumar and Chaudhari, Sachin and Varshney, Neeraj and Varshney, Pramod K},
  journal={IEEE Internet Things J.},
  volume={11},
  number={5},
  pages={8783--8799},
  month=may,
  year={2023},
}

@inproceedings{yin_joint_2025,
  title={Joint {Source} and {Channel} {Coding} for {Multi}-{Modal} {Satellite}-to-{Ground} {Semantic} {Communications}},
  author={Yin, Yanbo and Liu, Shu and Wen, Dingzhu and Wu, Youlong and Shi, Yuanming},
  booktitle={Proc. IEEE WCNC},
  month={Mar},
  year={2025},
  address={Milan, Italy},
  doi={10.1109/WCNC61545.2025.10978250}
}

@article{bui_semantic_2025,
  title={Semantic Image Encoding and Communication for Earth Observation With LEO Satellites},
  author={Bui, Van-Phuc and {Thinh Quang Dinh} and Leyva-Mayorga, Israel and Pandey, Shashi Raj and Lagunas, Eva and Popovski, Petar},
  journal={IEEE Trans. Cogn. Commun. Netw.},
  volume={11},
  number={2},
  pages={1210--1224},
  year={2025},
  month={Apr},
  doi={10.1109/TCCN.2024.3451724}
}

@article{chen_free_2025,
  title={Free Space Optical Semantic Communication for Satellite Remote Sensing Image Transmission},
  author={Chen, Wenbin and Ju, Cheng and Yuan, Tianxing and Zhan, Yueying and Zhang, Min and Wang, Danshi},
  journal={IEEE Trans. Commun.},
  year={2025},
  month={Apr},
  doi={10.1109/TCOMM.2025.3562356},
  note={{E}arly {A}ccess}}

@article{jiang_semantic_2025,
  title={Semantic {Satellite} {Communications} {Based} on {Generative} {Foundation} {Model}},
  author={Jiang, Peiwen and Wen, Chao-Kai and Li, Xiao and Jin, Shi and Li, Geoffrey Ye},
  journal={IEEE J. Sel. Areas Commun.},
  volume={43},
  number={7},
  pages={2431--2445},
  year={2025},
  month={Jul},
  doi={10.1109/jsac.2025.3559113},
}
